# Milled to order: toward predictive mechanochemistry of halide perovskites

*Susan A. Rigter,*[1] *Loreta A. Muscarella,*[2,*]

[1] Department of Physics & Astronomy, Vrije Universiteit Amsterdam, Amsterdam, The Netherlands

[2] Dipartimento di Fisica e Chimica – Emilio Segrè, University of Palermo, Palermo, Italy

*E-mail: loretaangela.muscarella@unipa.it

**Table of Content**

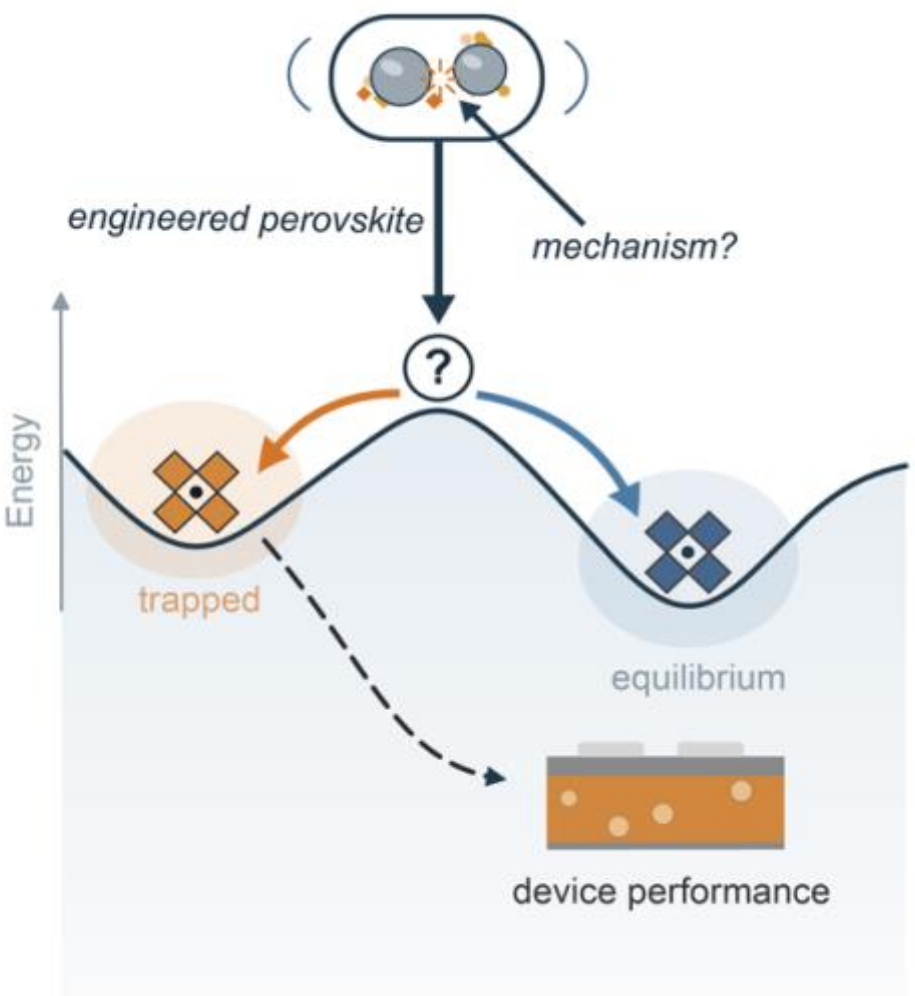

**Keywords**: halide perovskites, mechanochemistry, kinetic trapping, metastable phases, green metrics, solvent-free synthesis, material discovery

**Abstract**
Mechanochemical milling is regarded as a scalable, solvent-free method that makes halide perovskites that solution processing cannot. Its deeper promise, though, is that mechanical force could become a new axis of synthetic control to turn milling from an empirical method into a predictive science. Halide perovskites are the ideal platform to establish this: their soft lattices and low formation energies make them intrinsically responsive to mechanical activation. Realising this requires settling what milling actually makes, and how: can a milled phase genuinely be trapped outside equilibrium, or is it just a convenient way to the equilibrium phase, and by what atomistic pathway does it work? These questions are mostly answered by assumption, while the in-situ and computational tools that resolved them for other material classes already exist. Closing this gap will unlock two design capabilities: mechanical energy used as a synthetic variable to tune phase selection, and metastability made programmable, producing trapped phases, specific intermediates, and compositions that have no possible solution route. Finally, two tests decide whether that control reaches practice: the state engineered during milling must survive into a working device, and the method must prove genuinely greener across the full device life cycle, rather than just solvent-free.

## 1. Introduction

Mechanochemistry, one of the oldest chemical techniques possibly dating back to prehistoric times,[1] has proven to be a well-suited synthesis method for one of the newest classes of opto-electronic materials: halide perovskites. The rapid rise in optoelectronic applications of these materials has itself been the subject of extensive review, and so has their mechanochemical synthesis.[2] This technique is often justified on practical grounds: it is a simple technique that avoids toxic solvents of solution processing, and is easily scalable.[3–5] The case for mechanochemistry extends across energy materials broadly, and in 2019 IUPAC recognized it as an Emerging Technologie in Chemistry.[5–8] More material-specifically, mechanochemistry does not guarantee the same crystalline perfection as slow solution growth, but the well-documented defect tolerance of halide perovskites makes themforgiving of the resulting imperfections, and the method offers finer stoichiometric control.[9]

Yet, the appeal of mechanochemistry for halide perovskites extends further. These materials possess an unusual combination of properties that make them intrinsically responsive to mechanical activation. They have low formation energies and soft and highly anharmonic lattices that can accommodate substantial local distortions without catastrophic structural failure.[10–13] Additionally, they have exceptionally low migration barriers for ionic defects: halide ions and vacancies migrate on energy scales of only a few tenths of an electron volt.[14–17] In device operation, these properties underlie phase segregation,[18] and dynamic structural disorder.[12] These same characteristics give rise to a low mechanical energy barrier for material transformations, so that halide perovskites may be particularly susceptible to mechanically driven non-equilibrium processes, whose mechanistic origins remain unresolved. On these grounds, halide perovskites are not simply another material class to which mechanochemistry can be applied. They may be the material class in which mechanochemistry is most powerful, combining soft, low-barrier, defect tolerant lattices with a synthesis method that works by mechanical force.

This unusually strong pair of material and method is at the center of this Perspective: mechanical force could become a new axis of synthetic control for halide perovskites, as tunable as temperature or pressure in traditional synthesis methods, able to trap non-equilibrium phases by design and to exploit the vast $ABX_3$ compositional space[19] that other methods cannot reach. Realising this means milling will not just be a convenient synthesis route, but grows into a predictive method for materials design.

This ambition requires answering three questions. First, what does milling actually make, and how: does it genuinely traps non-equilibrium phases or does it merely offer a convenient route to equilibrium ones, and by what atomistic pathway do they form? Second, what would closing that gap unlock? We envision two capabilities: mechanical energy could be used as a synthetic variable, tuning phase selection through impact energy; and building on that, synthetic control for materials design and discovery, engineering known phases and intermediates, strain and defect states, trapped metastable phases, and compositions with no solution route at all. Third, does the engineered state survive from powder to film to a device, and if so, can a milled powder be turned into a good device, using a genuinely more sustainable method?

The question is no longer whether milling can make a halide perovskite, but whether grinding, the *oldest* way of making matter, can become a tool for designing one of the most exciting classes of *modern* materials while the mill is still turning.

## 2. The what and how of milling

The first question named in the Introduction has two parts, what milling makes, and how it makes it. Merely naming the phase resulting from milling, settles neither. The first part is whether a milled phase can genuinely be trapped outside equilibrium, or whether milling is only a convenient synthesis route for equilibrium products, a question of thermodynamics that needs independent reference. The second part is by what atomistic pathway a product forms, something a comparison of the end product can never reveal, and which matters because it sets the defect and strain state the phase carries. The same phase, made by two different milling pathways, can hold very different strain and defects, and it is that state, not just the phase label, that determines how the material performs in an eventual device. Answering both will close fundamental understanding gaps existing in halide perovskite mechanochemistry.

### 2.1 Kinetic trapping, or only compositional access?

What milling makes is easy to observe and easy to over-interpret. Producing a phase by milling does not by itself prove that mechanochemistry has accessed a non-equilibrium state, it could merely be that it has provided a convenient synthetic route to an equilibrium state. We distinguish compositional access from kinetic trapping. Compositional access refers to obtaining an equilibrium phase other methods cannot, while kinetic trapping, by contrast, refers to stabilizing a phase outside its equilibrium phase diagram. While the former can be established by showing that milling produces the intended, phase-pure, homogeneous material at ambient temperature and pressure, the latter cannot be inferred from the sample alone: it requires an external reference showing that this phase is not the equilibrium product under those conditions. While compositional access is valuable, allowing us to make materials that cannot be produced otherwise, if milling can kinetically trap too, this will extend the already vast compositional space of halide perovskites beyond the equilibrium phase diagram.

Whether that distinction can be made depends on the composition. It is useful to separate three situations (**Figure 1a**), increasing in difficulty to distinguish. In the first situation, milling produces a phase for which an independent thermodynamic reference, such as a phase diagram, is known and the milled phase can be compared to the known equilibrium state under the specific conditions. In the second situation, milling produces something for which no thermodynamic reference exists, but that can also be made using other methods: the two results can then be compared. In the last and most difficult case, milling provides compositional access, and reaches a composition that has not been realized otherwise. Only the first case allows kinetic trapping to be demonstrated rather than assumed, and it is much the rarest of the three in the literature.

Hong et al. provide a clear example: they reported several high-temperature perovskite phases, mechanochemically synthesized at room temperature, such as black-phase $CsPbI_3$.[20] This cubic phase is thermodynamically metastable at room-temperature, while the equilibrium is the yellow, orthorhombic phase (**Figure 1b**).[21] Reaching this phase using milling is strong evidence of kinetic trapping, and points precisely towards the mechanical responsiveness of halide perovskites we anticipate. To show that a mechanochemically trapped phase can be used in practice, Hong and colleagues went one step further, showing it survives into a working device, a topic this perspective will return to in part 4.1.[20] Yet, beyond this, true proof of kinetic trapping in mechanochemically synthesized halide perovskites remains scarce. It is, however, well-established in other material classes: for example, mechanical alloying has trapped equilibrium-forbidden metastable phases for decades.[22,23] Halide perovskites are a comparatively new system to which it appears this logic can be applied and possibly expanded.

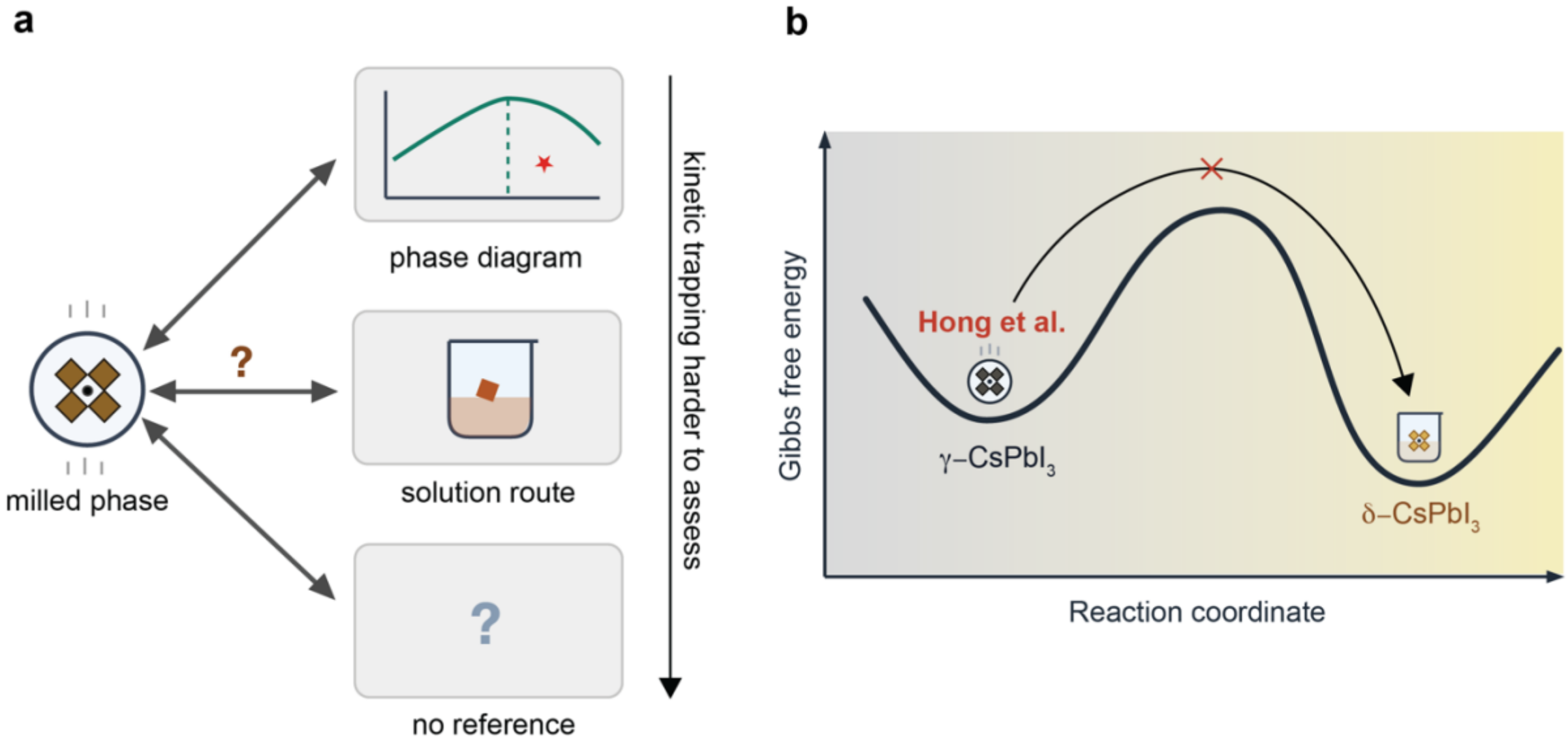

**Figure 1. Distinguishing kinetic trapping from compositional access in mechanochemical perovskite synthesis. a)** Conceptual framework for deciding whether a milled phase is kinetically trapped or a convenient route to the equilibrium product. Three situations are shown: an independent thermodynamic reference — a phase diagram — exists (top), so the milled phase can be compared with the known equilibrium and trapping can be proven; no phase diagram exists but the same composition is also accessible by a solution route (middle), allowing direct comparison that establishes route-equivalence but no equilibrium; neither a phase diagram nor an experimental counterpart exists (bottom), so compositional access and kinetic trapping cannot be distinguished. **b)** Gibbs free-energy landscape for $CsPbI_3$ illustrating the rare provable case. The metastable γ-$CsPbI_3$ perovskite (local minimum) is obtained directly by mechanosynthesis at room temperature,[20] whereas the thermodynamically stable δ-$CsPbI_3$ non-perovskite (global minimum, equilibrium at RT) is the product reached by conventional solution routes.

The second situation represents the bulk of the mechanochemical perovskite literature: it is also possible to make the resulting phase using other methods. Often, milling reproduces that same result it in a cleaner, more scalabe way. $CsPbBr_3$ and $MAPbI_3$ dominate the field, alongside families of well-known phases of 3D, 2D, and 0D halide perovskites.[24,25] Here, milling provides practical advantages. It can ease the route to compositions that are difficult to reach in solution, for instance fluorinated perovskites or double perovskites, where it avoids poor solubility of the precursors.[13,26–28] It provides exact stoichiometric control and phase-purity even for more complicated compositions,[29] and can combine kilogram-scale synthesis with powders that are shelf-stable for years as convenient feedstocks.[29–31]. Contrastingly, there are cases in which milling produces different phases of the same composition than other methods. On its own, this does not prove that milling produced an out-of-equilibruim phase, as, for example, a solution route does not by definition lead to the equilibrium product either.[32,33]

In the third situation, it is very complicated to understand whether a milled material is at equilibrium or not: it provides compositional access and reaches a material that has never been made before. With no phase diagram or experimental counterpart to compare against, there is no way to distinguish between compositional access or kinetic trapping, leaving only assumption. Examples are double perovskites, whose solution synthesis is prevented by the poor solubility of their precursors[9], and mixed-halide and mixed-cation compositions not achieved in solution.[34–36] In our view, the occurrence of these cases of compositional access will grow as halide perovskite milling matures. Mechanochemical milling is proven to be unusually good at making compositionally complex solids such as high entropy alloys that cannot be made phase-pure otherwise,[37–39] and the perovskite $ABX_3$ lattice with its tolerance for mixed site occupancy is an ideal host for such complexity, with true fine-tuning of their properties as the final goal.

Taken together, milling reaches perovskites no other method can, but the proof of whether it can kinetically trap halide perovskite phases is limited. In most cases, kinetic trapping cannot yet be distinguished from compositional access. Understanding this is an important ingredient for predictive mechanochemistry. However, even a phase that is proven to be trapped is only half of what will tell us what to target using predictive milling, as it tells us nothing on how it was formed. Awareness of the formation mechanism is a prerequisite for the ability to control a reaction, and it cannot be understood from the phase alone.

### 2.2 The Mechanistic Blind Spot

Whether or not milling can genuinely trap non-equilibrium phases, the atomistic route by which the final product forms is still unresolved. Yet, understanding it is the prerequisite of control over the synthesis. There are two possibly competing mechanisms. The first treats mechanochemical perovskite formation as a fully solid-state process,[40,41] in which cations and halides exchange through lattice diffusion and grain boundaries, driven by the mechanical energy from repeated impacts (**Figure 2a**). Via this route, phase trapping may arise because atomic rearrangements remain constrained by the precursor lattices. The second hypothesis is that impacts briefly drive the contact region into a highly disordered or partially molten state (**Figure 2b**). If these transient disordered states take place, phase selection may depend much more sensitively on impact energy, cooling rate, and local composition, instead causing trapping from a mechanically driven quench of a non-equilibrium intermediate. This idea is analogous with stress-induced local and virtual melting, which has been proposed to explain mechanically driven transformations in several non-perovskite systems.[42–45] This distinction mirrors metallurgy, where solid-state mechanical alloying and melt-quenching trap observably different metastable states for the same composition.[22] Resolving which mechanism operates, or if they are taking place simultaneously matters because they give rise to fundamentally different origins for kinetic trapping, and imply different control strategiesm which is fundamental to our vision in this perspective.

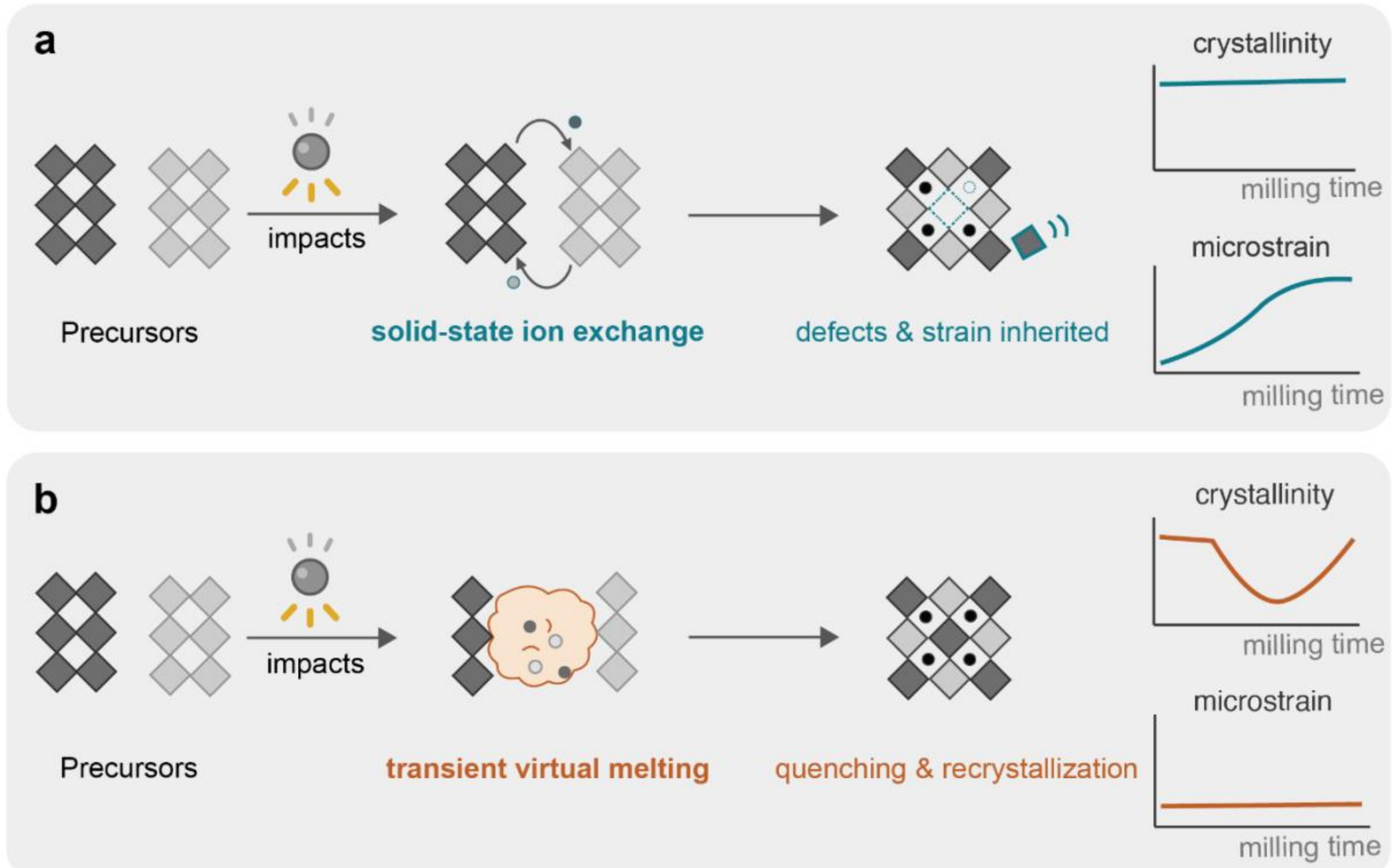

**Figure 2. Mechanistic models for mechanochemical halide perovskites formation. a)** In the solid-state ion-exchange mechanism, halide exchange occurs through diffusion across crystalline interfaces while the lattice remains intact. This process progressively accumulates defects and microstrain inherited from repeated mechanical impacts, with crystallinity largely preserved. **b)** In the transient virtual melting mechanism, local structural disorder enables rapid compositional homogenization before quenching and recrystallization into the final alloyed phase. This pathway predicts only limited residual microstrain but a transient loss of crystallinity during milling. The contrasting evolution of crystallinity and microstrain provides experimentally testable fingerprints to distinguish between these two fundamentally different reaction mechanisms.

An important other reason to unravel the formation mechanism, is that the two pathways could reach the same phase by opposite structural histories, but would arguably lead to fundamentally different defect populations. This, too, cannot be understood by only characterizing the final phase. Solid-state ion exchange forms the product without the reactants ever melting, so it inherits the defects and microstrain of the precursors, whereas a transient melt resolidifies into the final product with much of that history erased. The defect population can be inherited by the final device, even when the phase is not, and strongly affect its functioning.
Regarding the exact reactions of halide perovskite mechanochemistry, many studies do provide important insights, but remain inconclusive. Xiao et al. provided the first window to a lead halide perovskite during milling by tracking optical absorption and photoluminescence, and resolved real-time formation of a metastable phase. However, it is an optical measurement, not a continuous structural record of which lattice sites atoms occupy during the impact itself, nor on whether a disordered intermediate briefly forms.[17] Alternatively, structural monitoring of reactions has been done ex situ, for example by stopping the mill at intervals and measuring aliquots.[24,46] This leads to a limitation by construction: each data point requires measuring the material outside the environment where the reaction is happening. It can identify what intermediate milled phases the reaction proceeds along, but it cannot establish what is taking place inside the mill. Furthermore, measuring the bulk temperature during milling has provided the knowledge that it stays moderate, and originates from friction rather than exothermic reactions,[47] but a mass-averaged measurement cannot answer the open question. It demonstrates or rules out neither hypothesis, because any local disordering or melting would occur fast and locally. Unraveling the mechanism requires in-situ monitoring of the reaction fingerprint, i.e. the evolution of crystallinity and microstrain during milling, (**Figure 2**), with techniques that provide microscopic, local, and real-time detail.
The toolbox needed have matured on other materials than halide perovskites, such as metal–organic frameworks, pharmaceutical cocrystals, and single-component organic solids: probing the inside of the mill continuously with synchrotron X-ray diffraction at high time resolution, sometimes paired with in situ Raman spectroscopy for complementary vibrational information.[48–50] Alongside this, kinetic models developed, and thorough reviews have made real-time measurements that resolve the mechanism a standard part of the mechanochemistry toolbox. It has isolated the effect of a single milling parameter on a reaction, and identified short-lived reaction intermediates in real time.[51–54]
Despite that, this well-established method has been used only once for halide perovskites, on a relatively niche composition. Jöbsis and co-workers used in situ synchrotron X-ray diffraction,[48] resolving the full formation mechanism of double perovskites, elpasolites, during milling,[27] building on previous ex-situ mechanistic studies of the same material.[55] Their work yielded quantitative kinetics at rates comparable to solvent-based crystallization and the observation that microstrain, which is attributed to ions diffusing toward their final lattice positions, relaxes as the reaction approaches completion,[27] consistent with the observation that ions remain ionically mobile in this same material family after milling stops.[17] For this system, the continuous diffraction record looks point to solid state ion-exchange: crystalline phases interconverting, assembled by solid-state ion transport, although the authors note that possible amorphous fractions complicate a fully quantitative description.

Atomistic simulation is a more recent addition to the toolbox. It can supply exactly what the virtual melting hypothesis requires: collision-level insights into whether a typical impact supplies enough energy to cause a local melting point, or only to drive solid state ion hopping.[56] Meanwhile, the elementary steps on which the solid-state route depends, are thoroughly calculated using density functional theory, which has produced ion migration barriers of a few tenths of an eV and low defect formation energies across $MAPbI_3$ and related compositions.[14–16] These computations were done to explain device hysteresis and degradation rather than effects under the non-equilibrium conditions of milling, and although the conversion is not trivial, they contain the elements a solid-state cation- and halide-exchange model can be built from.

The two halves of the ‘what and how of milling’ have so far been answered separately, each on a different perovskite composition: kinetic trapping has been proven for an inorganic lead perovskite, and we know the full formation mechanism of a lead-free double perovskite. What predictive milling needs is a library of full pictures, answering both parts on the same, well-characterized halide perovskite composition. The more of these are established, the more reliably the field can extrapolate into the far larger perovskite space and predict the outcome of a synthesis a priori, rather than milling by trial and error, moving it toward deliberate material design.

## 3. From understanding to design

Closing the gaps mapped above turns the milling process from something described afterwards into something steered on purpose: once it is understood what milling does, the mechanical parameters can be controlled to reach chosen states rather than to find them accidentally. That control has two consequences. Mechanical energy can be used as a synthetic variable alongside the conventional temperature and pressure, to tune phase selection and material state through the energy and frequency of impact. In turn, this enables directed synthesis of known and unknown, stable and metastable phases, driving materials discovery and strain and defect engineering.

### 3.1 Mechanical energy as a new synthetic variable

Conventional synthesis explores equilibrium phase diagrams by tuning thermodynamic parameters such as temperature, pressure, and chemical potential, each of which moves the system predictably across known phase boundaries. Mechanochemistry adds a dimension of synthetic control this does not span: the energy landscape generated by repeated mechanical impacts, set by the energy and frequency of collisions, and by how that deposited energy dissipates into local stress and strain. If these quantities determine which transformations occur, then milling conditions become a form of synthetic control unavailable to solution or thermal routes: mechanical force as a parameter as tunable as temperature or pressure, but able to reach states that heat and pressure cannot, or only far from ambient conditions.

Establishing mechanical energy as a synthesis parameter for halide perovskites connects directly to both understanding gaps raised above. Understanding kinetic trapping determines how vast the accessible target space is, how far it can reach beyond equilibrium; and resolving the mechanism is what makes the process predictable at all, since the same delivered energy behaves differently if it drives solid-state ion hopping or instead depresses a local melting point.

The outline of such control is already visible in modelling literature, though not yet for halide perovskites. Gayen et al. built a molecular-dynamics protocol for a model ball-milling reaction and found that product formation is triggered only when the energy absorbed per ion pair in a single collision exceeds the crystal's cohesion energy. They show a universal quantitative collision-level threshold, independent of milling-ball shape, and stable across crystal sizes spanning nearly two orders of magnitude.[56] This is a simple form of the kind of rule that predictive mechanochemistry can be built from: one mapping a controllable input, impact energy, to a discrete outcome, whether the reaction proceeds.

This possibility is not just theory. Changing a single experimental milling parameter has already been shown to change the outcome. Real-time diffraction has isolated the effect of a single variable, the mass of the milling ball, on the transformation pathway of a molecular crystal, switching which route the reaction follows,[53] and mechanical alloying makes the same point quantitatively across metals, showing that milling intensity and time determine which phase forms.[23,41,57] There is even evidence of how sensitive this tuning can be for a halide perovskite: $CsPbBr_3$ reaches a phase-pure product within minutes of milling, far faster than the hours of high-energy milling required for metal alloying,[23,58] and overmilling rapidly degrades it. The reaction runs through three distinct stages: ternary intermediates, then a phase-pure perovskite, and finally material degradation, signalled by loss of photoluminescence.[24] These reactions quickly pass through reproducible intermediates rather than at random, which is the precondition for steering them.

The steps that remain are to assemble a large set of rules into a predictive map of milling conditions to outcomes, and then the demonstration of the newly found control: selecting a target material composition and state, and selecting the parameters to mill it on purpose.

### 3.2 Synthetic control for material design and discovery

With that control in hand, milling becomes more than a convenient synthesis tool: a way to make halide perovskites exactly to order, to design, engineer, and discover new materials. The possible targets range from known phases and intermediates, strain and defect engineering, trapping metastable phases, and ultimately the design and discovery of phases that have no other possible synthesis route.

Some of these possibilities are already emerging through the identification of mechanochemical intermediates. [22,23,41,59] For instance, $CsPb_2Br_5$ and $Cs_4PbBr_6$, the intermediates on the $CsPbBr_3$ pathway, are optoelectronic materials studied in their own right,[25] and the pathway to them is known. Similarly, the in situ study of $Cs_2AgBiBr_6$ revealed a previously unknown Bi-rich intermediate, $CsBi_2Br_7$, whose properties remain unexplored.[27] That same study also could rationalise why $In^{3+}$ and $Fe^{3+}$ enter the lattice by a different route than $Sb^{3+}$, and thus anticipate whether an alloyed intermediate should exist or not for a given composition.[27] Beyond identifying transient phases, resolved reaction pathways provide a predictive framework for synthesis. Once the key intermediate is known, precursor stoichiometry, milling sequence, or impact energy can be chosen to steer a reaction through it or around it, and the same framework could then be extended to compositions not yet explored.

Creating completely new perovskites is perhaps the most distinctive promise of predictive mechanochemistry: a discovery engine for compositions that have no other synthesis route at all. It is particularly powerful for compositionally complex systems, where solubility limits, competing phases, or kinetic barriers restrict conventional synthetic approaches. These include mixed-metal elpasolites, low-dimensional halide perovskites, mixed-halide systems, metal-free molecular perovskites,[60] and highly complex perovskite compositions, and are largely unexplored territories. Here, mechanical activation may reveal genuinely new states rather than reproduce known compounds.

That opportunity also ties back to the question about kinetic trapping. If kinetic trapping can be deliberately controlled, milling could provide access to metastable materials with no place on any known phase diagram, reachable only under the far-from-equilibrium conditions generated during mechanical activation. For metals, mechanical alloying has achieved this: milling parameters determine not only what metastable phase forms, but also how stable it is. We envision that an analogous predictive framework could ultimately emerge for halide perovskites. The control reaches beyond the final crystal phase. Because the formation pathway strongly influences the defect and strain state, selecting milling parameters can directly tailor that state. Rather than simply minimising defects, mechanochemistry could become a route to deliberately engineering defect populations[61] and strain states[62] optimised for targeted device functionalities. Ultimately, the target of predictive mechanochemistry need not be a crystal structure as such, but a desired functional property such as bandgap, luminescence, ionic conductivity, or defect landscape. The reaction pathway would then be engineered to produce the structural state required to achieve it.

Closing the understanding gaps outlined above will mature halide perovskite mechanochemistry from a convenient solvent-free synthesis route into a predictive synthetic platform, enabling rational control over reaction pathways, material states, and ultimately material functionality. For this platform to be meaningful, the engineered materials must reach a real device.

## 4. From material design to device

Everything discussed so far concerns the powder that comes straight out of the mill. Turning that powder into a working device requires an extra step, and that step determines whether the vision we have laid out matters in practice. Independent of the mechanism by which mechanochemical formation occurs, and in what way the material has been engineered, the powder must be converted into a high quality layer before it can be part of a functional device. In that conversion, there needs to be material memory: the structural, compositional, strain, and defect state created during milling must survive, rather than being erased when the powder is processed. That survival depends on the method used to turn the powder into a device layer, and is critical: it can be the point at which the previously described opportunities of mechanochemistry fall apart, and it is therefore important to investigate. If this survival is possible, two further questions must be answered: is it possible to make devices from milled powders that are of comparable quality to conventional perovskite device fabrication methods, and more sustainable than the routes it would replace?

### 4.1 What survives into a device

Detecting material memory in its simplest form, phase preservation, can be done with a routine measurement: a simple crystallographic measurement such as X-ray diffraction shows whether the material phase is preserved or reverted during processing steps such as dissolution, deposition, or annealing. This simple measurement applies to kinetically trapped materials, as well as to the far larger number of compositions and phases that are accessible only by milling. It becomes more nuanced when the milled material also has an engineered strain or defect state, although a lab that can assess this landscape in a powder often is also capable of assessing it in film. The more fundamental question, though, is not about whether it is possible for each material to determine whether the milled state is preserved after a device is made, but whether material memory can be generalized. Only then is it worth the investment to make mechanochemistry into the predictive platform described in Section 3.

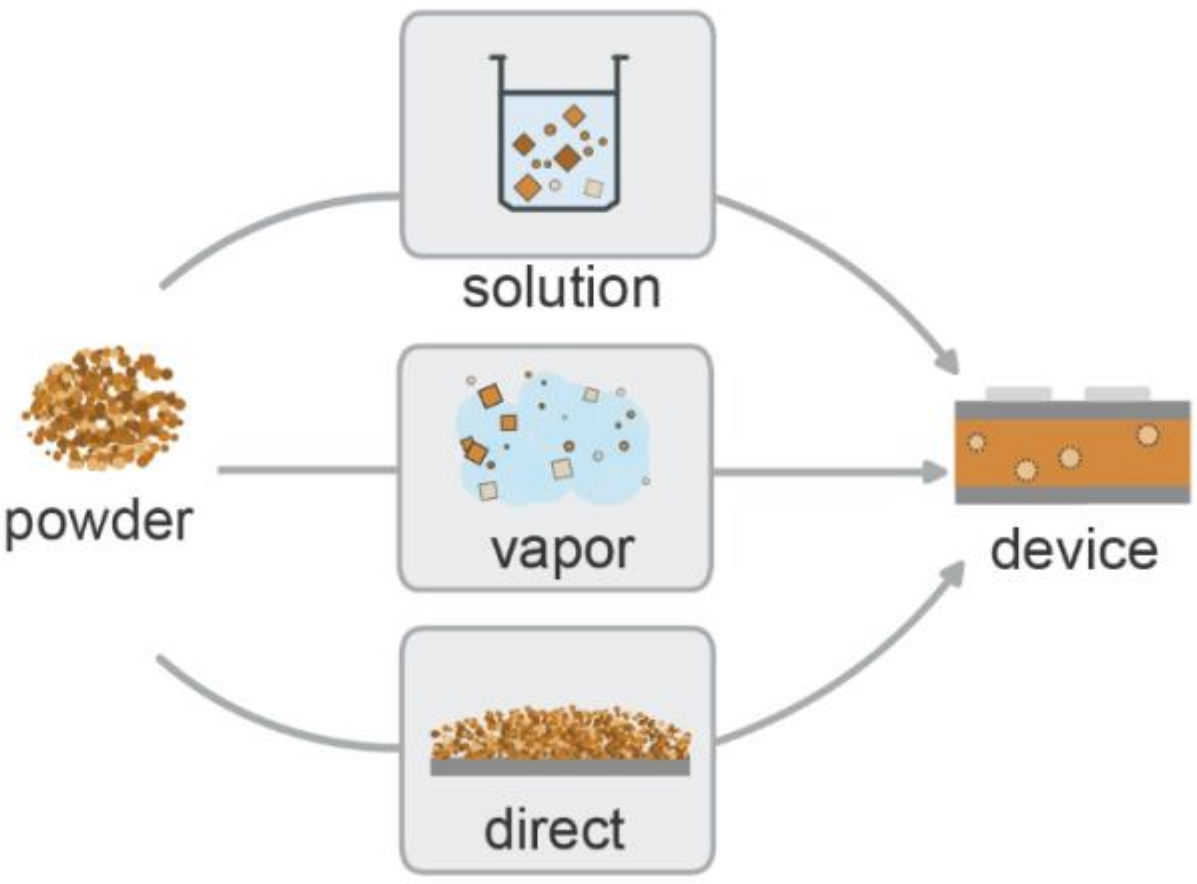

**Figure 3. From milled powder to device.** Three classes of deposition route convert a milled powder into a device layer: solution-based routes dissolve and recrystallize it, vapor-based routes avoid solvents but still recrystallize from the vapor, and direct routes such as pressing or powder aerosol deposition transfer the powder as a solid.

To understand what survives device fabrication, we have to categorize the deposition methods. Poli and Petrozza made such a categorization in a recent editorial, dividing perovskite film fabrication into solvent-based and solvent-free methods. This is a good starting point, as this distinction likely matters for cost and sustainability.[63] However, to properly consider material memory, we have to divide the solvent-free category further. While all solvent-based methods involve dissolving the milled powder and recrystallizing it, not all solvent-free methods require such a thorough disassembly. Besides solution-based methods, we therefore make the distinction between vapor-based methods, such as thermal evaporation or vacuum deposition that do depend on recrystallization, and direct deposition methods, such as pressing or aerosol deposition, which do not (**Figure 3**). The direct deposition methods are the most promising: the fewer changes the milled powder undergoes, the greater the likelihood that the engineered material retains all of its functional features. However, recrystallization does not necessarily erase the mechanochemical imprint, as there are indications that some of its features can survive even through these routes. Elucidating how the material evolves during deposition will require the in situ, mechanism-resolving measurements proposed above, now applied specifically to the deposition process.[63] Done systematically on a large set of well-characterized systems for each processing category, such measurements would show whether milled states survive processing methods.

Those studies do not yet exist. What does exist, are several ex-situ comparisons between powders and devices, but they are scattered across research groups, compositions, device architectures, and deposition routes. However, it is worth understanding what the scattered comparisons tell us about each processing route.

Solvent-based routes to make thin films are common in mechanochemical studies, as Palazon and coworkers highlighted. The milled powder is dissolved in solvents such as DMF or DMSO which is then, for example, spin coated, reintroducing a solvent-based processing step after the solvent-free synthesis.[9] Dissolution and recrystallization would seem likely to alter the carefully engineered milled material state. Counterintuively, though, starting the solvent-based process with a pre-made perovskite powder has advantages compared to directly dissolving precurors, which points to possible material memory. There are processing advantages: for a number of layered, double, and metal-free perovskite compositions, it is possible to dissolve and spincoat the milled powder, while it is impossible to create films of acceptable quality when starting directly from precursors.[64,65] Additionally, there are benefits even for compositions where spincoating precursors directly also works. Considering processing, a growing body of work shows that conventional precursor solutions degrade chemically with age, but redissolved pre-synthesized powders yield solutions seem largely insensitive to solution age.[31,66–69] Precursor inks retain significant structure in solution rather than dissolving into isolated ions, and therefore the synthetic history of a powder may persist into the final film rather than being completely erased. Also, the resulting films have higher crystallinity,[70] better stability, and lower defect density. [71,72]

However, there is some evidence that these advantages cannot be attributed to the exact powder material carrying over into the film: Zhang et al. fabricated high quality black-phase $FAPbI_3$ films by dissolving and spin coating yellow-phase milled powder, crediting the film quality to the powders stoichiometry instead.[72]

Next, we consider vapor-phase deposition, which includes techniques such as thermal evaporation, pulsed laser deposition, and vacuum deposition. Although it eliminates solution processing entirely, this route volatilizes and thoroughly recrystallizes the milled material, similar to the solvent-based route. For vapor methods, mechanochemistry offers a convenient single source,[73,74] a pre-made target that avoids delicately co-evaporating different components. However, it does not guarantee a film with the same phase and stoichiometry as the target, rendering full material memory unlikely. El Ajjouri et al.[75] found that the outcome of single-source vacuum deposition from mechanochemically synthesized inorganic halide perovskite powders is strongly halide dependent. Starting with a pure perovskite phase, the chloride composition deposited phase-pure, bromide retained phase impurities requiring high-temperature annealing, and iodide remained largely unreacted $PbI_2$ even after thermal treatment. When using pulsed laser-deposition, it is possible to stoichiometrically transfer the exact stoichiometry and phase of the initial target,[55] though often the stoichiometry of the initial target needs to be carefully engineered taking into account the volatility of the perovskite components.[76,77] In summary, while more thorough comparisons would be useful, the limited evidence implies that an engineered milled state is unlikely to survive into a film. However, vapor-phase methods can benefit from more controlled mechanochemistry for optimized target engineering.

For both the solvent-based and vapor-based routes, there is limited evidence of material memory, and to date, nothing shows that strain, defect states, or even phase survives the process, and little reason to expect it all would. The key unresolved question is whether the advantages of mechanochemistry in these routes originate from the powder itself, from the mechanical history encoded during synthesis, or from the ability to provide a highly controlled precursor for alternative deposition routes. Until more thorough comparisons between powders and resulting films resolve this, the more promising route to use for preservation of a milled stae, is direct deposition. This opportunity has been recognized since the earliest studies of mechanochemical perovskites. Hong et al. made a pellet of mechanochemically synthesized inorganic halide perovskite powders, like the ones used as targets in vapor deposition. They then used this pellet directly, applying electrodes to make a photodetector, precisely because dissolving or vaporizing the powder would partially erase the advantage of the synthesis route.[20] This is a promising method, since pellets have been built in a variety of working devices.[78,79]

However, these pressed pellets are thick, while perovskites only require a very thin film to make excellent devices. Powder aerosol deposition[80] offers a solution: dry powder aerosol is sprayed onto a substrate with high velocity fracturing larger particles into nanocrystallites on impact. This results in a compact, continuous layer with the same crystal and electronic structure as the initial powder.[81] The layer quality is good enough to use in a functional device, and can be further improved using thermal post-treatment, though this is likely to alter the strain and defect state of the material.[82]

Across all three routes, starting from a milled powder is beneficial. However, what is still unresolved is whether that benefit is real material memory, i.e. the milled state being carried through, or simply the advantage of starting with the exact desired mixture as precursor to a film. Settling this calls for a systematic comparison of powder and resulting films rather than the existing isolated demonstrations. Such a comparison would do more than answer that yes-or-no question. By tracking which structural, compositional, strain, and defect features survive each route, it would show that the routes differ less in whether they preserve the milled state than in which parts of it they preserve. Once preservation is resolved feature by feature rather than as a whole, it stops being a single goal to maximise. Ultimately, the best method for deposition is not always the one that preserves most of the material state, but depends on what aspects of the material were engineered, and the device being made. For example, preserving a phase or complex stoichiometry is a far looser requirement than keeping the exact strain and defect state. Before it is possible to predict which method suits each situation, rather than blindly choose one, we need better understanding of what aspects of the material are carried over in each route.

Ultimately, the significance of mechanochemical perovskite synthesis depends not only on whether milling makes new and engineered materials, but also on if these materials can be exploited in a device that is both good and genuinely greener due to milling.

### 4.2 Device quality and Sustainability

If the state of the milled material does carry over into a device, two further aspects of the device then decide whether this is worth the effort: is it possible to make a good quality device when starting with a milled powder, and is the process to make the device indeed more sustainable, a claim often made to legitimize the use of mechanochemistry, but rarely substantiated?

To our knowledge, the only studies comparing the quality of devices made directly from precursors and from milled powders, have done so via a solvent-based route. Prochowicz and co-workers used spin coating to make solar cells. A mixed-cation mechanoperovskite solar cell reached a considerably higher efficiency than its precursor-derived analogue, because the former was more phase-pure.[83] Yet in a later, more optimized study, the two routes resulted in similar efficiency devices. However, the films of mechanochemical origin showed better stoichiometric control, lower hysteresis and an order-of-magnitude reduction in interfacial trap density, highlighting benefits of mechanochemistry regardless.[71]

Contrastingly, Leupold and coworkers fabricated solar cells based on $MAPbI_3$, where observed slightly inferior device metrics from mechanochemical powders. This was attributed to the absence of residual $PbI_2$, a beneficial accident of solution-processing that can passivate defects at interfaces despite being an impurity from a structural perspective.[30] Combining this with the result of Prochowicz's systems, it is clear that although mechanochemically derived films are of higher quality, this does not always result in higher efficiency device. The evidence for whether milling results in a better quality device, points in different directions, and it is unclear if milling just results in a convenient feedstock for solution processing, or whether there are more fundamental advantages. Perhaps the process can be optimized, and mechanochemistry is indeed the key to better quality devices across the board.

Fortunately, the device-level case for mechanochemistry does not depend on whether the resulting devices are strictly higher efficiency than counterparts originating from different methods. When making new perovskite materials using the mechanochemical control envisioned above, device efficiency will follow from optimally engineered materials, as long as it is possible to find a method to retain the engineered state and fabricate a good quality device. For each processing route, the resulting device quality is encouraging. For solvent-based approaches to processing mechanochemical powders, by far the most explored route, the performance ceiling for mechanochemically derived devices is nearly as good as the state of the art. For example, $FAPbI_3$ solar cells have reached efficiencies of over 24%[72,84], approaching record efficiency for this composition.[85] Powder-derived devices are therefore no longer technologically immature or inferior. Good effiency devices have also been made from single source pulsed laser deposition,[76] and the first signs of well-working devices from direct deposition methods are arising.[82,86] It is likely that well-functioning devices can be made through either of the three processing routes, and selection depends most strongly on the required preservation of the material state.

Another thing that route selection depends on in practice, is sustainability. Mechanochemistry is often promised to be more sustainable than other routes, but this only holds up if the whole device fabrication remains cleaner. This requires adding up the overall footprint of mechanochemistry: precursor synthesis, milling energy, scale-up, and downstream film processing, and comparing it with other device fabrication methods, which to date has not been done for many halide perovskites. Doing so requires quantitative green metrics, and those already exist: Fantozzi et al. set out a systematic framework for atom economy, E-factor, process mass intensity, reaction mass efficiency, and life-cycle indicators, with dozens of quantified comparisons against solution routes across amide bonds, peptides, ureas, hydantoins, porphyrins, and esters, while noting that such calculations are still not undertaken systematically across mechanochemistry. [87]

This halide perovskite gap is not for lack of applicability. Ceriotti et al. compute exactly this comparison for fluoride perovskites $KCuF_3$ and $KNiF_3$, and the result is stark: atom economy of 100% and process mass intensity of 1 for the mechanochemical route, against 51.7% and 191 for $KCuF_3$ or 37.6% and 294 for $KNiF_3$ by a mild solvothermal route, at roughly a quarter of the energy.[26] This preliminary result is encouraging, but not sufficient. Besides applying these methods to more compositions, the "solvent-free" label of mechanochemistry also needs to be revisited: although most milling recipes remove solvents from the initial synthesis, the most used route to fabricate devices reintroduces it, and all three possible routes add to the energy demand; quantifying the milling process alone is not sufficient. Whether milling is quantifiably greener than the routes it replaces is a different question, one that demands a full accounting of milling energy, materials and subsequent processing, and for lead halide perovskites that accounting is still largely absent. The contrast is sharpened by how thoroughly the solution-processing method is quantified: there are harmonized life-cycle assessments of solution-processed perovskite modules,[88] a detailed analysis of the health and environmental burden of the most common solvents,[89] and green-solvent substitutions, e.g. replacing DMF and DMSO with biomass-derived γ-valerolactone, have been reported to halve manufacturing costs and cut climate-change impact by 80%.[90] Solution processing has therefore established the baseline against which alternatives can be measured; mechanochemical synthesis has not done so for even a single lead halide composition.

## 5. Summary and Outlook

We have argued that mechanochemistry offers halide perovskites much more than a solvent-free, scalable route to known phases. Its unrealised promise is to use mechanical force as a synthetic variable, used to target types of halide perovskites other methods cannot, such as kinetically trapped phases and highly complex compositions. Reaching it runs through a chain of questions this Perspective has tried to untangle: what milling actually makes and by what mechanism, what that understanding would unlock, and whether the engineered state survives into a device that is both good and genuinely greener. None of these is yet settled for a halide perovskite, but none is out of reach, either.

What they share is a striking feature: the tools to answer these questions already exist. Four toolboxes would have to be imported from other fields: real-time measurement from molecular mechanochemistry, parameter–outcome mapping from mechanical alloying, collision-level simulation from computational chemistry, and quantitative life-cycle metrics from green chemistry. The fifth is homegrown: the halide perovskite community has a deep predictive understanding of ion migration and defect energetics, built to explain how finished devices operate and degrade, which could be aimed at how the material forms, instead.

The obstacle to a predictive mechanochemistry of halide perovskites is therefore not invention but integration of these toolboxes, existing across communities that have so far studied diverging problems. No material system is better suited to this: halide perovskites and mechanochemistry are an unusually strong pair. Soft, anharmonic lattices, exceptionally low ion-migration barriers, and rich metastable behaviour make these materials primed to respond to mechanical force, and make them a sensitive testbed for the underlying questions of mechanochemical reactivity, metastability, and defect evolution.

Should this all succeed, the oldest way of making matter turns into a way of predictively designing modern, functional materials.

**Acknowledgement**

L.A.M., and S.A.R. received funding from the Dutch Research Council (NWO) under the grant numbers OCENW.XS25.3.003 and OCENW.XS26.1.278.

**Conflict of Interests**

The authors declare no conflict of interests.